\documentstyle[12pt,twoside]{article}
\begin{document}

\newif\iffigs\figsfalse
\figstrue

\iffigs
  \input epsf
\else
  \message{No figures will be included. See TeX file for more
information.}
\fi

\thispagestyle{empty}
\begin{flushright}
MPI-PhT/97-38
\end{flushright}
\bigskip\bigskip\bigskip\begin{center}
{\LARGE {Application of a coordinate space method}}
\vskip 7pt
{\LARGE {for the evaluation of lattice Feynman diagrams}}
\vskip 7pt
{\LARGE {in two dimensions}}
\end{center}  
\vskip 1.0truecm
\centerline{Dong-Shin Shin}
\vskip5mm
\centerline{Max-Planck-Institut f\"ur
 Physik}
\centerline{ -- Werner-Heisenberg-Institut -- }
\centerline{F\"ohringer Ring 6, 80805 Munich, Germany}
\vskip 2cm
\bigskip \nopagebreak \begin{abstract}
\noindent
We apply a new coordinate space method for the evaluation of lattice
Feynman diagrams suggested by L\"uscher and Weisz to field theories
in two dimensions.
Our work is to be presented for the theories with massless propagators.
The main idea is to deal with the integrals
in position space by making use of the recursion relation 
for the free propagator $G(x)$ which allows to compute 
the propagator recursively by its values around origin.
It turns out that the method is very efficient 
and gives very precise results.
We illustrate the technique by evaluating a number of 
two- and three-loop diagrams explicitly.
\end{abstract}
\vskip 1.5cm

\newpage\setcounter{page}1

\def\eqb{\begin{equation}}
\def\eqe{\end{equation}}
\def\dib{\begin{displaymath}}
\def\die{\end{displaymath}}
\def\eqnb{\begin{eqnarray}}
\def\eqne{\end{eqnarray}}
\def\eqsb{\begin{eqnarray*}}
\def\eqse{\end{eqnarray*}}
\def\md{\mbox{d}}
\def\me{\mbox{e}}

\section{Introduction}

It is generally believed that QCD
describes the strong interaction of quarks and gluons.
QCD is, however, very complicated to investigate.
Therefore, it is useful to study at first 
similar, but simpler models 
and apply the related knowledge gained to the real situation.
The most frequently studied toy models for this 3+1 dimensional theory
are the 1+1 dimensional models,
e.g. the non-linear O($n$) $\sigma$ model in two dimensions,
which have also found wide applications 
in the framework of the lattice field theory.
They have not only the simple theoretical structure, 
but also are, due to low dimensionality, simple to simulate numerically.
Another high interest in these two-dimensional models 
is to be found in perturbation theory 
which plays an important conceptual and practical role 
in lattice field theory.
However, the evaluation of lattice Feynman diagrams is, 
in general, not easy.
On the lattice the integrands become non-trivial
function of internal and external momenta, which makes it difficult
to apply the standard tools in continuum perturbation theory.

A few years ago, L\"uscher and Weisz \cite{LuWe}
suggested a new position space method 
for the evaluation of lattice Feynman diagrams in four dimensional
Yang-Mills theories,
inspired by the observation of Vohwinkel~\cite{Vo} that
the free lattice propagator can be calculated recursively
by its values around origin. 
Their method concerns a technique of computing Feynman integrals 
in the coordinate space
which allows to obtain very accurate results
with little effort and small amount of computer time.
In our work on the computation of four-loop $\beta$ function 
in the two-dimensional non-linear O($n$) $\sigma$ model \cite{Sh},
we have made extensive use of this technique.
Since we have found it very useful,
we would like, in the present paper, to
describe the method in the case of two dimensions.
This will be done by choosing some typical diagrams and evaluating them
explicitly.

To illustrate the basic ideas to be discussed in the present paper,
we take the following 3-loop integral into consideration
which will be treated in detail in the main text:\footnote{Throughout 
the whole paper, we will work on a two-dimensional square lattice $\Lambda$
and keep the lattice constant $a$ being equal to 1.}
\eqb\label{latsumv}
B_1 = \int^{\pi}_{-\pi}\frac{\md^2q}{(2\pi)^2}
\frac{\md^2r}{(2\pi)^2}\frac{\md^2s}{(2\pi)^2}
\frac{\sum_{\mu=0}^1\hat q_{\mu}\hat r_{\mu}\hat s_{\mu}\hat t_{\mu}}
{\hat q^2\hat r^2\hat s^2\hat t^2},\hspace{1cm}t=-q-r-s,
\eqe
where
\eqb
\hat p_{\mu}=2\sin\frac{p_{\mu}}{2}, \hspace{1cm}
\hat p^2=\sum_{\mu=0}^1\hat p_{\mu}^2
\eqe
and $p=(p_0,p_1)$ denotes the line momentum.

There are a few known conventional 
computer programs available to compute this integral.
The numerical accuracy which can be achieved by these methods
is, however, strongly limited and thus they are not very useful
if one requires to have results with high precision.
Considering the integral in position space, however,
provides us with a new possibility to compute it very accurately with
high efficiency.

In position space the integral (\ref{latsumv})
has the form
\eqb\label{b1sum}
B_1 = \sum_{x\in\Lambda}\sum_{\mu=0}^1\Big[\partial_{\mu}G(x)\Big]^4,
\eqe
where $\partial_{\mu}$ denote the lattice derivatives and 
$G(x)$ is the position space free massless propagator
\eqb\label{freeprg}
G(x) = \int^{\pi}_{-\pi}\frac{\md^2p}{(2\pi)^2}\frac{\me^{ipx}-1}{\hat p^2}.
\eqe

The free propagator
$G(x)$ diverges logarithmically at large $x$.
After differentiation,
$\partial_{\mu}G(x)$ falls off like $|x|^{-1}$ and 
the sum (\ref{b1sum}) is therefore absolutely convergent. 
In principle, one can then compute $B_1$ 
by summing over all points of $x$ up to some finite value, e.g. $|x|\leq 30$. 
However, since the convergence is slow, 
one requires to sum over large range of $x$ 
if one would like to have precise result.
Taking the limited computing availability into account, 
this is, of course, not the most efficient way.
By studying the large $x$ behavior of the propagator $G(x)$,
it is however possible to improve on the accuracy systematically 
until the desired level of precision is reached.
It is the aim of the present paper to show how this can be done.

In next section we discuss the properties of 
the free propagator $G(x)$ and, in particular,
derive a recursion relation
which allows us to express the propagator
as a linear combination of 
its values near origin.
These results will then be applied to the computation of
lattice sums in section~\ref{secets} where
we take various examples to illustrate the procedure.
It turns out that the method is quite efficient
and gives very accurate results.
Finally, in section~\ref{secpdp} 
we present a little more subtle case of evaluating 
Feynman diagrams with non-zero external momenta, particularly in the 
continuum limit. 
The usefulness of the position space method in this case as well
is demonstrated by dealing with a number of diagrams explicitly.
We conclude the paper by including some detailed computations
in a series of appendices to make it more readable.

\section{\mbox{Investigation of the free propagator} 
\mbox{and zeta functions}}
\setcounter{equation}{0}

In this section we prepare for the main work in
the next two sections
where we show how the position space technique 
can be applied to the evaluation of Feynman diagrams.
This requires the study of large $x$ behavior of the free propagator.
We also need to derive the recursion relation for it
which is essential to the computation of the integrals.

\subsection{Properties of the propagator $G(x)$}

The propagator $G(x)$ is a Green function for $\triangle$, i.e. it satisfies
the laplace equation
\eqb\label{lapla}
-\triangle G(x) = \delta^{(2)}(x),
\eqe
where $\delta^{(2)}(x)$ is the lattice $\delta$-function and 
$\triangle$ denotes the lattice laplacian
\eqb
\triangle = \sum_{\mu=0}^1 \partial_{\mu}^*\partial_{\mu}.
\eqe
Here,
$\partial_{\mu}$ and $\partial_{\mu}^*$ are 
the forward and backward lattice derivatives respectively:
\eqnb
\partial_{\mu}f(x) &=& f(x+\hat{\mu}) - f(x), \\
\partial_{\mu}^*f(x) &=& f(x) - f(x-\hat{\mu}).
\eqne
We note the relation 
$\partial_\mu^* = -\partial_\mu^\dagger$ for the backward lattice derivative,
where $\partial_\mu^\dagger$
is the adjoint of the forward lattice derivative defined with
respect to the inner product 
$(\partial_\mu^\dagger f,g)=(f,\partial_\mu g)$.

A rigorous derivation of the large $x$ behavior of $G(x)$ shows that
in the limit $x\to\infty$ the propagator
diverges logarithmically and has the form
\eqb\label{gxcm}
G^{{\rm c}}(x) = -\frac{1}{4\pi}(\ln x^2 + 2\gamma+3\ln 2)
\eqe
which corresponds to the leading behavior in the continuum limit.
By making use of the laplace equation (\ref{lapla}),
one can also work out the subleading terms systematically
to obtain the asymptotic expansion. Up to order $|x|^{-6}$, we find
\eqb\label{gxlx}
G(x) =
G^{{\rm c}}(x)
-\frac{1}{4\pi}\bigg[\frac{1}{(x^2)^3}Q_1
+ \frac{1}{(x^2)^6}Q_2
+ \frac{1}{(x^2)^9}Q_3
+ \cdots \bigg],
\eqe
where
\eqnb\label{defq0}
Q_1 &=& \frac12(x^2)^2 -\frac23x^4, \\
Q_2 &=& -\frac{371}{120}(x^2)^4+\frac{47}{5}x^4(x^2)^2-\frac{20}{3}(x^4)^2, \\
Q_3 &=& \frac{4523}{56}(x^2)^6 - \frac{7657}{21}x^4(x^2)^4 
+ \frac{3716}{7}(x^4)^2(x^2)^2 - \frac{2240}{9}(x^4)^3.
\eqne
In the above expansions we have introduced the shorthand notation
\eqb\label{shonot}
x^n = \sum_{\mu=0}^1 (x_{\mu})^n.
\eqe
We remark that 
in two dimensions the general relation 
\eqb\label{gereca}
x^{2n+2}=x^2x^{2n}-\frac12x^{2n-2}\Big[(x^2)^2-x^4\Big]\,, 
\hspace{0.7cm} n\geq 1
\eqe
is valid
so that the expansion terms can always be expressed
by the powers $x^2$ and $x^4$ 
since higher powers are to be reduced to these two powers
(note $x^0=2$).
For example, from Eq.~(\ref{gereca}) we find the identities
\eqnb\label{identx6}
x^6 &=& \frac12\Big[3x^2x^4-(x^2)^3\Big], \\
x^8 &=& \frac12\Big[(x^4)^2+2(x^2)^2x^4-(x^2)^4\Big], \\
x^{10} &=& \frac14x^2\Big[5(x^4)^2-(x^2)^4\Big] .
\eqne
The detailed discussion of the asymptotic behavior of $G(x)$ is
presented in Appendix~\ref{ap1}.

Now we would like to derive a recursion formula for the free propagator
$G(x)$ which makes it possible to determine $G(x)$ for large $x$ by
its initial values around origin.
The key relation for obtaining the formula is the identity,
first observed by Vohwinkel \cite{Vo},
\eqb\label{recuhx}
G(x+\hat{\mu}) - G(x-\hat{\mu}) = x_{\mu}H(x),
\eqe
where $H(x)$ is independent of $\mu$ and has the form
\eqb
H(x) = \int_{-\pi}^{\pi}\frac{\md^2p}{(2\pi)^2}
\me^{ipx}\ln\hat p^2.
\eqe
By summing over $\mu$ and using Eq. (\ref{lapla}) for $x\not=0$,
one can eliminate $G(x+\hat{\mu})$ in Eq. (\ref{recuhx})
and express $H(x)$ like
\eqb\label{fhx}
H(x) = \frac{2}{\sum_{\mu=0}^1x_{\mu}}
\Big[2G(x)-\sum_{\mu=0}^1G(x-\hat{\mu})\Big].
\eqe
If one inserts this formula into Eq. (\ref{recuhx}),
one finally obtains the recursion relation
\eqb\label{gxrecu}
G(x+\hat{\mu}) = 
\frac{2x_{\mu}}{\sum_{\mu=0}^1x_{\mu}}
\Big[2G(x)-\sum_{\mu=0}^1G(x-\hat{\mu})\Big] +
G(x-\hat{\mu})
\eqe
which is valid for $x\not=0$.

Since the propagator has the properties
\eqnb
G(x_0,x_1)&=&G(x_1,x_0),\\
G(x_0,x_1)&=&G(-x_0,x_1),
\eqne
$G(x)$ is completely characterized by $x_0\geq x_1\geq 0$.
In this sector, Eq. (\ref{gxrecu}) is a recursion formula
which allows us to express $G(x)$ as a linear combination of 
three initial values
\eqnb\label{iniv00}
G(0,0) &=& 0, \\
G(1,0) &=& -\frac14, \\\label{iniv11}
G(1,1) &=& -\frac{1}{\pi},
\eqne
i.e. the values of the propagator at the corners of the unit square.
Therefore, the propagator has the general form
\eqb\label{gxcan}
G(x) = -\frac14r_1(x) -\frac{1}{\pi}r_2(x),
\eqe
where $r_1(x)$ and $r_2(x)$ are rational numbers 
which are to be computed recursively by Eq. (\ref{gxrecu}).

In this place we would like to stress the fact that the analytic knowledge 
of the initial values (\ref{iniv00})--(\ref{iniv11})
is crucial to the precise determination of the propagator $G(x)$
for large $x$. The reason for this is that 
$r_1(x)$ and $r_2(x)$ increase exponentially with $x$
while $G(x)$ grows only logarithmically
and thus there is a large cancellation of terms 
on the r.h.s. of Eq.~(\ref{gxcan}).

\subsection{Properties of the function $G_2(x)$}

Another function which often appears 
in the procedure of evaluating diagrams is
\eqb\label{defg2}
G_2(x) = \int^{\pi}_{-\pi}
\frac{\md^2p}{(2\pi)^2}\frac{\me^{ipx}-1+
\frac12\sum_{\mu=0}^1\hat p_{\mu}^2x_{\mu}^2}{(\hat p^2)^2}.
\eqe
The laplacian of this function satisfies
\eqb\label{lapg2}
\triangle G_2(x) = -G(x).
\eqe

One can also compute the function $G_2(x)$ analytically 
in the limit of large $x$ and expand it systematically by using the expansion
(\ref{gxlx}) and the laplace equation (\ref{lapg2}).
Up to order $|x|^{-4}$, we find in this way
\eqb\label{laxg2}
G_2(x) =
\frac{1}{16\pi}\bigg[x^2R_1 + R_2 + \frac{1}{(x^2)^5}R_3 + 
\frac{1}{(x^2)^8}R_4 + \cdots\bigg],
\eqe
where
\eqnb
R_1 &=& \ln x^2 + 2\gamma+3\ln 2-2, \\
R_2 &=& 
-\frac12\ln x^2+\frac{1}{3}\frac{x^4}{(x^2)^2}-\frac12(2\gamma+3\ln 2), \\
R_3 &=& \frac{19}{120}(x^2)^4 - \frac{19}{15}x^4(x^2)^2
+ \frac{4}{3}(x^4)^2, \\ 
R_4 &=& - \frac{10183}{1680}(x^2)^6
+ \frac{14041}{420}x^4(x^2)^4
- \frac{1214}{21}(x^4)^2(x^2)^2
+ \frac{280}{9}(x^4)^3.
\eqne
For the explicit derivation of the large $x$ expansion we refer to
Appendix~\ref{ap2}.

Similarly to the propagator $G(x)$,
one can also derive a recursion relation for $G_2(x)$.
The first step for the derivation is the observation
\eqb\label{helpsfg2}
G_2(x+\hat{\mu}) - G_2(x-\hat{\mu}) = -x_{\mu}H_2(x),
\eqe
where $H_2(x)$ is independent of $\mu$ and has the form
\eqb
H_2(x) = G(x)+\frac{1}{4\pi}.
\eqe
Eliminating $G_2(x+\hat{\mu})$ in Eq. (\ref{helpsfg2}) 
by using Eq. (\ref{lapg2}) gives the formula
\eqb
H_2(x) = \frac{1}{\sum_{\mu=0}^1x_{\mu}}
\Big[G(x)-4G_2(x) + 2\sum_{\mu=0}^1G_2(x-\hat{\mu})\Big]
\eqe
which leads straightforwardly to the recursion relation 
\eqb\label{recrelg2}
G_2(x+\hat{\mu}) = 
-\frac{x_{\mu}}{\sum_{\mu=0}^1x_{\mu}}
\Big[G(x)-4G_2(x) + 2\sum_{\mu=0}^1G_2(x-\hat{\mu})\Big]+G_2(x-\hat{\mu})
\eqe
for $x\not=0$.

Like the propagator $G(x)$, due to the properties
\eqnb
G_2(x_0,x_1)&=&G_2(x_1,x_0),\\
G_2(x_0,x_1)&=&G_2(-x_0,x_1),
\eqne
$G_2(x)$ is completely characterized by $x_0\geq x_1\geq 0$.
In this sector, the recursion relation (\ref{recrelg2}) 
allows $G_2(x)$ to be expressed in terms of 
$G(x)$ and a linear combination of 
its values at the corners of the unit square,
\eqb
G_2(0,0)\,,\hspace{0.2cm}G_2(1,0)\,,\hspace{0.2cm}G_2(1,1)\,.
\eqe
These initial values can also be computed analytically with the results
\eqnb
G_2(0,0) &=& 0\,, \\
G_2(1,0) &=& 0\,, \\
G_2(1,1) &=& \frac{1}{8\pi}\,.
\eqne

\subsection{Zeta functions $Z(s,h)$}

In the procedure of evaluating the lattice sum like Eq. (\ref{b1sum}),
zeta functions appear.
As a preparation for the computation of such lattice sums in next section,
we discuss the evaluation of the
zeta functions $Z(s,h)$ associated with the square lattice $\Lambda$.

Let us define a genealized zeta function through
\eqb\label{zetaf}
Z(s,h) = {\sum_{x\in\Lambda}}'h(x)(x^2)^{-s},
\eqe
where 
$h(x)$ is a harmonic homogeneous polynomial in the real variables
$x=(x_0,x_1)$ with even degree $d$ and 
$s$ complex number with $\mbox{Re}\,s>1+d/2$.
The primed summation symbol implies that the point $x=0$ should be 
omitted in the summation.
Note that the sum (\ref{zetaf}) is absolutely convergent 
in the specified range of $s$.

The zeta functions can be evaluated as follows. We first introduce 
the heat kernel
\eqb\label{heatk}
k(t,h) = \sum_{x\in\Lambda}h(x)\me^{-\pi tx^2}.
\eqe
In terms of the heat kernel, 
Eq. (\ref{zetaf}) can be rewritten in the form
\eqb\label{zetahk}
Z(s,h) = \frac{\pi^s}{\Gamma(s)}\int_0^{\infty}\md t\,t^{s-1}[k(t,h)-h(0)].
\eqe
Now we observe that the heat kernel has the property
\eqb\label{hkident}
k(t,h) = (-1)^{d/2}t^{-d-1}k(1/t,h)
\eqe
which follows from the poisson summation formula
\eqb\label{poissonfom}
\sum_{x\in\Lambda}\me^{-iqx}\me^{-\pi tx^2}
=t^{-1}\sum_{x\in\Lambda}\me^{-(q+2\pi x)^2/(4\pi t)}
\eqe 
and the condition that $h(x)$ is a harmonic function.

If we divide the integral (\ref{zetahk}) into two parts and 
make use of the identity (\ref{hkident}), 
we finally obtain the representation 
\eqb\label{zetasol}
Z(s,h) = \frac{\pi^s}{\Gamma(s)}\bigg\{\frac{h(0)}{s(s-1)}
+ \int_1^{\infty}\md t\Big[t^{s-1}+(-1)^{d/2}t^{d-s}\Big]\Big[k(t,h)-h(0)\Big]
\bigg\}.
\eqe
In the integration range $1\leq t<\infty$,
the large $x$ terms in the series (\ref{heatk}) 
are exponentially suppressed.
Eq. (\ref{zetasol}) can therefore be computed 
numerically very precisely with little effort, 
e.g. by using symbolic language MATHEMATICA.

\section{Evaluation of lattice sums}\label{secets}
\setcounter{equation}{0}

\begin{figure}[htb]
\begin{center}
\leavevmode
\epsfxsize=60mm
\epsfbox{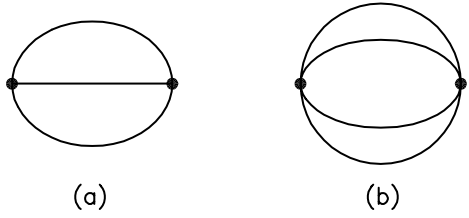}
\end{center}
\caption{Two- and three-loop diagrams}
\label{scalar23}
\end{figure}

By dealing with a number of integrals explicitly, 
we would now like to show 
how the results of the last section are to be applied to the evaluation of 
lattice Feynman diagrams. 
Concretely, we will consider in the following
the two- and three-loop diagrams of figure~\ref{scalar23}.

\subsection{Two-loop Feynman integrals}

As a first example
let us consider the integral,
encountered in the 3-loop $\beta$ function computation 
in the O($n$) $\sigma$ model,
\eqb\label{anal124}
A_1 = \int^{\pi}_{-\pi}
\frac{\md^2k}{(2\pi)^2}\frac{\md^2l}{(2\pi)^2}
\frac{1}{\hat k^2\hat l^2\hat{s}^2}
\sum_{\mu=0}^1\hat k_{\mu}\hat l_{\mu}\hat s_{\mu}^2
\cos\frac{s_{\mu}}{2}\,,\hspace{0.8cm}
s=k+l
\eqe
which corresponds to the first diagram in figure~\ref{scalar23}.

In position space this integral can be written as
\eqb\label{a1cons}
A_1 = \frac14\sum_{x\in\Lambda}\tilde A_1(x),
\eqe
where
\eqb\label{a1pos}
\tilde A_1(x) = \sum_{\mu=0}^1
(\partial_{\mu}-\partial_{\mu}^*)G(x)
\Big\{\Big[(\partial_{\mu}-\partial_{\mu}^*)G(x)\Big]^2 +
\Big[(\partial_{\mu}+\partial_{\mu}^*)G(x)\Big]^2\Big\}.
\eqe
$G(x)$ diverges logarithmically at large $x$ and
an explicit computation of large $x$ behavior of Eq.~(\ref{a1pos})
shows that $\tilde A_1(x)$ falls off like $|x|^{-4}$.
The sum (\ref{a1cons}) is therefore absolutely
convergent and a first approximation to $A_1$ may be obtained 
by summing over all points $x$ with (say) $|x|\leq 30$.
A direct summation in this way gives a result as a first approximation
\eqb
A_1 = -0.04166(1).
\eqe

We can now improve on the accuracy by investigating the 
asymptotic behavior of $\tilde A_1(x)$.
From the asymptotic form of $G(x)$ [Eq. (\ref{gxlx})], 
we deduce that 
\eqb
\tilde A_1(x) =\Big\{\tilde A_1(x)\Big\}_{\infty}+{\cal O}(|x|^{-8}),
\eqe
where
\eqb\label{lxea1}
\Big\{\tilde A_1(x)\Big\}_{\infty}=
-\frac{1}{2\pi^3}\bigg\{\frac{1}{(x^2)^2}
-2\frac{x^4}{(x^2)^4}
-\frac{10}{(x^2)^3}
+32\frac{x^4}{(x^2)^5}
-24\frac{(x^4)^2}{(x^2)^7}\bigg\}.
\eqe
Taking the asymptotic expansion of Eq.~(\ref{lxea1})
into consideration, $A_1$ can be rewritten as
\eqb
A_1=\frac14\Big[A^{'}_1+A^{''}_1+\tilde A_1(0)\Big],
\eqe
where
\eqnb
& & A'_1 = {\sum_{x\in\Lambda}}'
\Big[\tilde A_1(x)-\Big\{\tilde A_1(x)\Big\}_{\infty}\Big], \\
& & A^{''}_1 = {\sum_{x\in\Lambda}}'\Big\{\tilde A_1(x)\Big\}_{\infty}, \\
& & \tilde A_1(0) = -\frac14 .
\eqne

The sum $A'_1$ where the asymptotic series has been subtracted
converges much more rapidly than the original sum $A_1$.
This subtracted sum can therefore be computed very accurately 
by summing over all points $x$ with $|x|\leq 30$. 
As for the term $A^{''}_1$ from the asymptotic expansion,
it can be computed with help of zeta functions very accurately as well.
For the purpose of applying the zeta functions, 
we transform the asymptotic expansion (\ref{lxea1})
in the following way:
\eqb
\Big\{\tilde A_1(x)\Big\}_{\infty}=
\frac{1}{(2\pi)^3}\bigg\{
\bigg[\frac{2}{(x^2)^2}+\frac{1}{(x^2)^3}\bigg]h_0(x)+
\bigg[\frac{2}{(x^2)^4}+\frac{4}{(x^2)^5}\bigg]h_1(x)+
\frac{3}{(x^2)^7}h_2(x)\bigg\}.
\eqe
Here, $h_0(x)$, $h_1(x)$ and $h_2(x)$ are 
the harmonic homogeneous polynomials given by
\eqnb\label{harmh0}
h_0(x) &=& 1,\\
h_1(x) &=& 4x^4-3(x^2)^2,\\\label{harmh2}
h_2(x) &=& 32(x^4)^2-48(x^2)^2x^4+17(x^2)^4.
\eqne
In Appendix~\ref{aphhp} we discuss how one computes the harmonic polynomials.
The subtraction term $A^{''}_1$ 
can then be written as a sum of zeta functions:
\eqb
A^{''}_1 = 
\frac{1}{(2\pi)^3}\Big\{2Z(2,h_0)+Z(3,h_0)+
2Z(4,h_1)+4Z(5,h_1)+3Z(7,h_2)\Big\}
\eqe
to which now the solution (\ref{zetasol}) can be applied.

Summing all three contributions together, we get finally
\eqb
A_1 = -0.041666666666(1).
\eqe
It seems very likely that $A_1$ is $-\frac{1}{24}$.
Indeed, the analytic computation of $A_1$ confirms this hypothesis 
(see Appendix~\ref{ap3}).
We note that it is straightforward to determine $A_1$ more precisely 
just by including more terms in the large $x$ expansion of (\ref{lxea1}).

It is interesting to compare our determination with 
the earlier estimation by Weisz \cite{We}
who states the value
\eqb
A_1 = -0.0416666(2).
\eqe
We clearly see 
that the result from our method is much more precise than that of Weisz.

As a second example, we consider the following integral 
which also appears in the 3-loop $\beta$ function computation
(see figure~\ref{scalar23})
\eqb
A_2 = \int^{\pi}_{-\pi}
\frac{\md^2k}{(2\pi)^2}\frac{\md^2l}{(2\pi)^2}
\frac{\hat{s}^2-\hat k^2-\hat l^2}{\hat k^2\hat l^2}
\frac{\hat{s}^4}{(\hat{s}^2)^2}, \hspace{1cm}s=k+l.
\eqe
Here, we made use of the notation of Eq.~(\ref{shonot}).

The evaluation of this integral proceeds in exactly the same way as
that of $A_1$.
In terms of the propagator $G(x)$ and the function $G_2(x)$ 
[Eq. (\ref{defg2})], in the position space
the integral can be expressed by
\eqb
A_2 = -\frac12\sum_{x\in\Lambda}\tilde A_2(x),
\eqe
where
\eqb
\tilde A_2(x) = \bigg\{\sum_{\mu=0}^1
\partial_{\mu}^*\partial_{\mu}\partial_{\mu}^*\partial_{\mu}G_2(x)\bigg\}
\bigg\{\sum_{\mu=0}^1\Big[
\Big((\partial_{\mu}-\partial_{\mu}^*)G(x)\Big)^2+
\Big((\partial_{\mu}+\partial_{\mu}^*)G(x)\Big)^2\Big]\bigg\}.
\eqe
The terms in this sum fall off like $|x|^{-4}$ and a direct evaluation
without subtraction gives
\eqb
A_2 = -0.18465(1).
\eqe

The convergence can be improved  by subtracting the asymptotic form 
of $\tilde A_2(x)$
which can be deduced from the asymptotic forms of $G(x)$ and $G_2(x)$.
Up to order $|x|^{-6}$, we find for the large $x$ expansion of $\tilde A_2(x)$
\eqb
\tilde A_2(x) =\Big\{\tilde A_2(x)\Big\}_{\infty}+{\cal O}(|x|^{-8}),
\eqe
where
\eqb
\Big\{\tilde A_2(x)\Big\}_{\infty}=
\frac{1}{16\pi^3}\bigg\{\frac{5}{(x^2)^3}h_0(x)
+\bigg[\frac{8}{(x^2)^4}+\frac{2}{(x^2)^5}\bigg]h_1(x)
+\frac{25}{(x^2)^7}h_2(x)\bigg\}.
\eqe

The subtracted sum 
\eqb
A'_2 = {\sum_{x\in\Lambda}}'
\Big[\tilde A_2(x)-\Big\{\tilde A_2(x)\Big\}_{\infty}\Big]
\eqe
is now rapidly convergent and may therefore be computed accurately.
On the other hand, the subtraction term
\eqb
A^{''}_2 = {\sum_{x\in\Lambda}}'
\Big\{\tilde A_2(x)\Big\}_{\infty}
\eqe
can be expressed as a sum of zeta functions:
\eqb
A^{''}_2 = 
\frac{1}{16\pi^3}\bigg\{5Z(3,h_0)+8Z(4,h_1)+2Z(5,h_1)+25Z(7,h_2)
\bigg\}
\eqe
which can be computed very accurately, too.

Since
\eqb
A_2=-\frac12\Big[A^{'}_2+A^{''}_2+\tilde A_2(0)\Big],
\eqe
we obtain finally
\eqb
A_2 = -0.1846545169(1).
\eqe
This result is also to be compared with the
earlier estimation by Weisz \cite{We} who states
\eqb
A_2 = -0.1846544(1).
\eqe

\subsection{Three-loop Feynman integrals}

Now we come to the 3-loop integral of Eq.~(\ref{latsumv})
which we have already discussed in Introduction
(see figure~\ref{scalar23}).
The summand of $B_1$ from Eq. (\ref{b1sum})
\eqb\label{tilb1}
\tilde B_1(x) = \sum_{\mu=0}^1\Big[\partial_{\mu}G(x)\Big]^4
\eqe
falls off like $|x|^{-4}$ at large $x$ and a direct summation
without subtraction gives
\eqb
B_1 = 0.0169611(1).
\eqe

For the acceleration of the convergence, we expand Eq.~(\ref{tilb1})
for large $x$. Up to order $|x|^{-6}$, we find
\eqb
\tilde B_1(x) =
\Big\{\tilde B_1(x)\Big\}_{\infty}+{\cal O}(|x|^{-8}),
\eqe
where
\eqb
\Big\{\tilde B_1(x)\Big\}_{\infty}=
\frac{1}{32\pi^4}
\bigg\{
2\frac{x^4}{(x^2)^4}
-\frac{3}{(x^2)^3}
-6\frac{x^4}{(x^2)^5}
+16\frac{(x^4)^2}{(x^2)^7}
\bigg\}.
\eqe
From the identity
\eqb
B_1 = B_1^{'} + B_1^{''} + \tilde B_1(0)
\eqe
with
\eqnb
& & B_1' = {\sum_{x\in\Lambda}}'
\Big[\tilde B_1(x)-\Big\{\tilde B_1(x)\Big\}_{\infty}\Big], \\
& & B_1'' = {\sum_{x\in\Lambda}}'\Big\{\tilde B_1(x)\Big\}_{\infty}, \\
& & \tilde B_1(0) = \frac{1}{128}
\eqne
and the zeta function representation
\eqb
B_1'' = 
\frac{1}{64\pi^4}\Big\{3Z(2,h_0)+4Z(3,h_0)+Z(4,h_1)+9Z(5,h_1)+Z(7,h_2)\Big\},
\eqe
it is straightforward to compute $B_1$ to get a more precise result
\eqb
B_1 = 0.016961078576(1).
\eqe
We remark that this integral has been computed by
Caracciolo and Pelissetto \cite{CaPe} who state the value
\eqb
B_1 = 0.016961.
\eqe

As another example,
we consider the 3-loop integral of the form
\eqb
B_2 = \int^{\pi}_{-\pi}\frac{d^2q}{(2\pi)^2}
\frac{d^2r}{(2\pi)^2}\frac{d^2s}{(2\pi)^2}
\frac{\Big(\sum_{\mu=0}^1\hat q_{\mu}\hat r_{\mu}\hat s_{\mu}\hat t_{\mu}
\Big)^2}{\hat q^2\hat r^2\hat s^2\hat t^2},\hspace{1cm}
t=-q-r-s.
\eqe
In terms of the position space propagator $G(x)$,
the integral can be expressed by
\eqb
B_2 = \sum_{x\in\Lambda}\tilde B_2(x),
\eqe
where
\eqb
\tilde B_2(x) = \sum_{\mu=0}^1\sum_{\nu=0}^1
\Big[\partial_{\mu}\partial_{\nu}G(x)\Big]^4.
\eqe
The terms in this sum fall off like $|x|^{-8}$ and 
converge very well.
Therefore, the direct summation already gives a quite precise result:
\eqb
B_2 = 0.13661977236(1).
\eqe

Through exactly the same procedure as $B_1$,
we obtain the following identity for $B_2$
which should give more accurate result:
\eqb\label{b2sum}
B_2 = {\sum_{x\in\Lambda}}'
\Big[\tilde B_2(x)-\Big\{\tilde B_2(x)\Big\}_{\infty}\Big]
+\frac{1}{32\pi^4}\Big\{3Z(4,h_0)+Z(8,h_2)\Big\} + \tilde B_2(0)
\eqe
with
\eqnb
& & \Big\{\tilde B_2(x)\Big\}_{\infty} =
\frac{1}{8\pi^4}\bigg\{\frac{5}{(x^2)^4}
-12\frac{x^4}{(x^2)^6}+8\frac{(x^4)^2}{(x^2)^8}\bigg\}, \\
& & \tilde B_2(0) = \frac{34}{\pi^4}-\frac{36}{\pi^3}+\frac{15}{\pi^2}
-\frac{3}{\pi}+\frac14 .
\eqne
The numerical evaluation of Eq.~(\ref{b2sum}) then 
yields a more precise value
\eqb
B_2 = 0.136619772367581(1).
\eqe
This number can also be compared with that by 
Caracciolo and Pelissetto \cite{CaPe}:
\eqb
B_2 = 0.1366198.
\eqe

Finally, we consider the 3-loop integral of the form
\eqnb
B_3 &=& \int^{\pi}_{-\pi}\frac{\md^2q}{(2\pi)^2}
\frac{\md^2r}{(2\pi)^2}\frac{\md^2s}{(2\pi)^2}
\frac{\Big[\widehat{(q+r)}^2-\hat q^2-\hat r^2\Big]
\Big[\widehat{(s+t)}^2-\hat s^2-\hat t^2\Big]}
{\hat q^2\hat r^2\hat s^2\hat t^2},\nonumber\\ & &
\hspace{4cm}t=-q-r-s.
\eqne
The position space representation of this integral is given by
\eqb\label{sumb3}
B_3 = \frac14\sum_{x\in\Lambda}\tilde B_3(x),
\eqe
where
\eqb
\tilde B_3(x) = \bigg\{\sum_{\mu=0}^1\Big[
\Big(
(\partial_{\mu}-\partial_{\mu}^*)G(x)\Big)^2+
\Big(
(\partial_{\mu}+\partial_{\mu}^*)G(x)
\Big)^2\Big]\bigg\}^2.
\eqe

The evaluation of the sum (\ref{sumb3}) runs in exactly the same way as above
and we get the result
\eqb
B_3 = 0.09588764425(1),
\eqe
where we have evaluated the sum by subtracting 
the asymptotic terms from the summand 
in Eq.~(\ref{sumb3}).
Finally, this result is to be compared with that in Ref.~\cite{CaPe}:
\eqb
B_3 = 0.095887.
\eqe

\section{Evaluation of diagrams with external momenta}\label{secpdp}
\setcounter{equation}{0}

\begin{figure}[htb]
\begin{center}
\leavevmode
\epsfxsize=110mm
\epsfbox{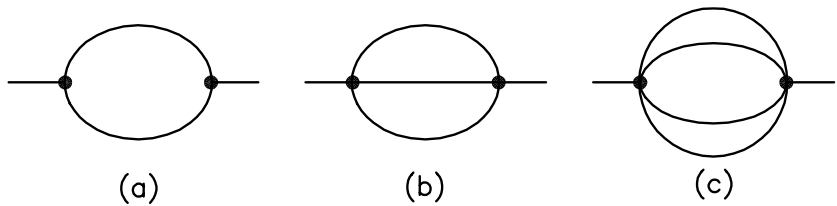}
\end{center}
\caption{One-, two- and three-loop diagrams with non-zero external momenta}
\label{pdep123}
\end{figure}

In this section, we consider the Feynman diagrams in figure~\ref{pdep123}
with external momenta~$p$.
We would like to evaluate the integrals in the continuum limit
which corresponds to taking $p$ to zero
since we are using lattice units and there are no mass parameters.
In other words, we have to work out
the leading terms of the integrals 
in an asymptotic expansion for $p\to 0$.

For the evaluation of diagrams with external momenta $p$,
it is necessary to solve the $p$-dependent zeta functions.
As a preparation for the actual treatment of such diagrams, 
we therefore discuss at first 
the zeta functions which are depending on the external momenta.

\subsection{Zeta functions ${\cal Z}(p,s)$}

Let us consider the $p$-dependent zeta functions of the kind ($s\geq1$)
\eqb\label{pdzeta}
{\cal Z}(p,s)={\sum_{x\in\Lambda}}'\frac{\mbox{e}^{-ipx}}{(x^2)^s}.
\eqe
With help of the heat kernel [Eq. (\ref{heatk})],
we can rewrite Eq. (\ref{pdzeta}) in the form
\eqb
{\cal Z}(p,s) = \frac{\pi^s}{\Gamma(s)}\int_0^{\infty}\md t\,t^{s-1}
\Big(\sum_{x\in\Lambda}\me^{-ipx}\me^{-\pi tx^2}-1\Big).
\eqe

We are interested in the evaluation of this integral in the limit $p\to 0$.
For this purpose, we consider the integral by
dividing the integration range into two parts
\eqb
{\cal Z}(p,s) = {\cal Z}'(p,s)+{\cal Z}''(p,s),
\eqe
where
\eqnb
{\cal Z}'(p,s) &=& \frac{\pi^s}{\Gamma(s)}\int_1^{\infty}\md t\,t^{s-1}
\Big(\sum_{x\in\Lambda}\me^{-ipx}\me^{-\pi tx^2}-1\Big), \\
{\cal Z}''(p,s) &=& \frac{\pi^s}{\Gamma(s)}\int_0^{1}\md t\,t^{s-1}
\Big(\sum_{x\in\Lambda}\me^{-ipx}\me^{-\pi tx^2}-1\Big).
\eqne
Now, note that ${\cal Z}'(p,s)$ is well-defined for all $s\geq 1$. 
We can therefore set $p$ to zero in that function.
Then, ${\cal Z}'(0,s)$ can be evaluated numerically very accurately
since the large $x$ terms in the series of the function
are exponentially suppressed.

On the other hand, ${\cal Z}''(p,s)$ contains singular term for small $p$.
For the evaluation of this function,
we first apply the Poisson summation
formula [Eq.~(\ref{poissonfom})] to rewrite
\eqb\label{serint}
{\cal Z}''(p,s) = \frac{\pi^s}{\Gamma(s)}\int_0^{1}\md t\,t^{s-1}
\Big[t^{-1}\sum_{x\in\Lambda}\me^{-(p+2\pi x)^2/(4\pi t)}-1\Big].
\eqe
In the series of the integrals appearing in Eq.~(\ref{serint}),
the only dangerous term for $p\to 0$ is that with $x=0$.
Hence, we split the integral ${\cal Z}''(p,s)$
into the regular and singular parts
\eqb
{\cal Z}''(p,s) = {\cal Z}''_r(p,s) + {\cal Z}''_s(p,s),
\eqe
where 
\eqnb
{\cal Z}''_r(p,s) &=& \frac{\pi^s}{\Gamma(s)}{\sum_{x\in\Lambda}}'
\int_1^{\infty}\md t\,t^{-s}\me^{-(p+2\pi x)^2t/(4\pi)}, \\
{\cal Z}''_s(p,s) &=& \frac{\pi^s}{\Gamma(s)}\int_0^{1}\md t\,t^{s-1}
[t^{-1}\me^{-p^2/(4\pi t)}-1] .
\eqne
After having set $p$ to zero in the well-defined function 
${\cal Z}''_r(p,s)$,
the integral ${\cal Z}''_r(0,s)$ can also be calculated numerically 
with high precision.
As for ${\cal Z}''_s(p,s)$, this function can easily be evaluated analytically
in the continuum limit
by appropriate separation of the singularity.

Summarizing the whole terms, we get finally for $p\to 0$
\eqnb
\lefteqn{{\cal Z}(p,s) = \frac{\pi^s}{\Gamma(s)}\bigg\{
{\sum_{x\in\Lambda}}'\int_1^{\infty}\md t\,t^{s-1}\me^{-\pi x^2t}+
{\sum_{x\in\Lambda}}'\int_1^{\infty}\md t\,t^{-s}\me^{-\pi x^2t}-\frac1s}
\nonumber\\\label{pzetasol} & &
+\rho^{1-s}\bigg[
\int_0^1\md z\,z^{s-2}\me^{-1/z}+
\int_1^{\rho}\md z\,z^{s-2}\Big(\me^{-1/z}-1\Big)+
\int_1^{\rho}\md z\,z^{s-2}\bigg]\bigg\}
\eqne
with $\rho = 4\pi/p^2$.

\subsection{Scalar one-loop integral}

We consider the integral
\eqb\label{onelint}
D_1(p) = \int^{\pi}_{-\pi}\frac{\md^2k}{(2\pi)^2}
\frac{\hat p^2-\hat k^2-\hat s^2}{\hat k^2\hat s^2}\,, \hspace{0.8cm} s = k+p
\eqe
which corresponds to the first diagram in figure~\ref{pdep123}.
By applying the coordinate space method,
we would like to show that $D_1(p)$ has the form 
\eqb\label{apform}
D_1(p) = c_1\ln p^2 + c_2 + {\cal O}(p^2)
\eqe
and determine the coefficients $c_1$ and $c_2$.\footnote{In this section 
${\cal O}(p^n)$ stands for 
a remainder $R(p)$ such that $\lim_{p\to 0}R(p)/|p|^{n-\epsilon}=0$
for all $\epsilon>0$.}

In position space, Eq. (\ref{onelint}) has the form
\eqb
D_1(p) = -\sum_{x\in\Lambda}\me^{-ipx}\triangle G(x)^2.
\eqe
For large $x$, the function $\triangle G(x)^2$ has the asymptotic behavior
\eqb
\triangle G(x)^2=\Big\{\triangle G(x)^2 \Big\}_{\infty}+{\cal O}(|x|^{-6})
\eqe
with
\eqb
\Big\{\triangle G(x)^2 \Big\}_{\infty} =
\frac{1}{2\pi^2}\frac{1}{x^2}
\bigg\{1-\frac72\frac{1}{x^2}
+5\frac{x^4}{(x^2)^3}\bigg\}.
\eqe
This suggests the splitting of $D_1(p)$ into four parts:
\eqb
D_1(p) = -\Big[D_1'(p) + D_1''(p) + D_1'''(p) + \triangle G(x)^2|_{x=0}\Big],
\eqe
where 
\eqnb
& & D_1'(p) = {\sum_{x\in\Lambda}}'\me^{-ipx}
\Big[\triangle G(x)^2 - \Big\{\triangle G(x)^2 \Big\}_{\infty}\Big], \\
& & D_1''(p) = {\sum_{x\in\Lambda}}'\me^{-ipx}
\Big[
\Big\{\triangle G(x)^2 \Big\}_{\infty}-\frac{1}{2\pi^2}\frac{1}{x^2}\Big], \\
& & D_1'''(p) = \frac{1}{2\pi^2}
{\sum_{x\in\Lambda}}'\frac{\me^{-ipx}}{x^2}, \\
& & \triangle G(x)^2|_{x=0} = \frac14 .
\eqne

The functions $D_1'(p)$ and $D_1''(p)$ are well-defined
and therefore the limit $p\to 0$ exists.
After having set $p$ to zero,
$D_1'(0)$ can be computed numerically with high precision
by applying the position space method. We receive the result
\eqb
D_1'(0)=-0.14215315037(1).
\eqe
$D_1''(0)$ can also be computed very accurately by making use of
the solution of zeta functions [Eq. (\ref{zetasol})]. 
A straightforward calculation gives
\eqb
D_1''(0)=0.275883297410294(1).
\eqe
The function $D_1'''(p)$, however, diverges in the continuum limit,
which makes it neces\-sary to apply Eq. (\ref{pzetasol}).
After some algebra and numerical evaluations, we find
\eqb
D_1'''(p)= - \frac{1}{2\pi}\ln p^2 + 0.1678588533415551(1).
\eqe

Summing all four contributions together, 
we confirm that $D_1(p)$ has indeed
the form of Eq. (\ref{apform}), and for the constants $c_1$ and $c_2$
we obtain
\eqnb
c_1 &=& \frac{1}{2\pi}, \\
c_2 &=& -0.55158900038(1).
\eqne

\subsection{Scalar two-loop integral}

Now let us consider an example of two-loop diagrams.
Concretely, we would like to discuss the following integral shown in 
the second diagram of figure~\ref{pdep123}:
\eqb\label{d2pem}
D_2(p) = \int^{\pi}_{-\pi}
\frac{\mbox{d}^2k}{(2\pi)^2}\frac{\md^2l}{(2\pi)^2}\frac{\md^2s}{(2\pi)^2}
\frac{\widehat{(k+l)}^2-\hat k^2-\hat l^2}{\hat k^2\hat l^2}
\frac{1}{\hat s^2}
\Big(\bar\delta_{p+k+l+s,0}-\bar\delta_{p+k+l,0}\Big).
\eqe
We have used here the shorthand notation 
$\bar\delta_{p,0}=(2\pi)^2\delta^{(2)}(p)$.

In position space, the integral (\ref{d2pem}) can be written in the form
\eqb\label{a2p}
D_2(p) = -\frac12\sum_{x\in\Lambda}\mbox{e}^{ipx}\tilde D_2(x)
\eqe
with
\eqb\label{tilax}
\tilde D_2(x) =
G(x)\sum_{\mu=0}^1\Big\{
\Big[(\partial_{\mu}-\partial_{\mu}^*)G(x)\Big]^2+
\Big[(\partial_{\mu}+\partial_{\mu}^*)G(x)\Big]^2\Big\}.
\eqe
For the evaluation of the sum (\ref{a2p}), we first 
observe the expansion terms of $\tilde D_2(x)$ for large $x$.
Up to order $|x|^{-4}$, we find
\eqb
\tilde D_2(x) = \Big\{\tilde D_2(x)\Big\}_{\infty} + {\cal O}(|x|^{-6}),
\eqe
where
\eqnb
\Big\{\tilde D_2(x)\Big\}_{\infty} &=&
-\frac{1}{4\pi^3}\bigg\{\frac{\ln x^2}{x^2}+\frac{2\gamma+3\ln2}{x^2}
\nonumber\\ & & 
-\frac72\frac{1}{(x^2)^2}\bigg[\ln x^2+3\ln 2+2\gamma-\frac{1}{7}\bigg]
\nonumber\\ & & 
+5\frac{x^4}{(x^2)^4}\bigg[\ln x^2+2\gamma+3\ln 2-\frac{2}{15}\bigg]\bigg\}.
\eqne
Now we separate the singular part in Eq. (\ref{a2p}) and rewrite it as
\eqb
D_2(p) = -\frac12\Big[D'_2(p) + D''_2(p) + D'''_2(p)\Big],
\eqe
where
\eqnb
D'_2(p) &=& {\sum_{x\in\Lambda}}'\mbox{e}^{ipx}
\Big[\tilde D_2(x)-\Big\{\tilde D_2(x)\Big\}_{\infty}\Big], \\
D''_2(p) &=& {\sum_{x\in\Lambda}}'\mbox{e}^{ipx}
\bigg[\Big\{\tilde D_2(x)\Big\}_{\infty}
+\frac{1}{4\pi^3}\bigg(\frac{\ln x^2}{x^2}+\frac{2\gamma+3\ln2}{x^2}\bigg)
\bigg], \\
D'''_2(p) &=& -\frac{1}{4\pi^3}{\sum_{x\in\Lambda}}'\mbox{e}^{ipx}
\bigg(\frac{\ln x^2}{x^2}+\frac{2\gamma+3\ln2}{x^2}\bigg).
\eqne
We note that $\tilde D_2(0)=0$. 

The function $D'_2(p)$ is well-defined in the limit $p\to 0$ and
can thus be computed numerically in this limit
by applying the position space method.
We obtain 
\eqb
D'_2(0) = 0.0705019591(1).
\eqe
$D''_2(0)$ can also be evaluated very accurately by using 
the solution of zeta functions [Eq.~(\ref{zetasol})]:
\eqb
D''_2(0) = -0.1385014044234234(1).
\eqe
The function $D'''_2(p)$, however, diverges in the continuum limit.
To this divergent function we can again
apply the solution (\ref{pzetasol}) as in the case of the one-loop integral.
The only difference here is the appearence of a logarithmic term in zeta 
function which can, however, 
be evaluated by differentiating the appropriate zeta 
function with respect to $s$.
In this way we obtain
\eqb
D'''_2(p)= -\frac{1}{8\pi^2}(\ln p^2)^2+\frac{5\ln 2}{(2\pi)^2}\ln p^2
-0.0841257673993860(1).
\eqe

Summing all three terms together, we finally get for $D_2(p)$
\eqb
D_2(p) = \frac{1}{(4\pi)^2}(\ln p^2)^2-\frac{5\ln 2}{8\pi^2}\ln p^2
+ \kappa_1
\eqe
with $\kappa_1$ given by
\eqb
\kappa_1 = 0.0760626064(1).
\eqe

\subsection{Scalar three-loop integral}

As the last example, we consider the 3-loop integral of the form
(see figure~\ref{pdep123})
\eqnb
D_3(p) &=& \int^{\pi}_{-\pi}
\frac{\mbox{d}^2k}{(2\pi)^2}\frac{\md^2l}{(2\pi)^2}\frac{\md^2s}{(2\pi)^2}
\frac{\md^2t}{(2\pi)^2}
\frac{\widehat{(k+l)}^2-\hat k^2-\hat l^2}
{\hat k^2\hat l^2\hat s^2\hat t^2} \nonumber\\ & &
\cdot\Big(\bar\delta_{p+k+l+s+t,0}-\bar\delta_{p+k+l+s,0}-
\bar\delta_{p+k+l+t,0}+\bar\delta_{p+k+l,0}\Big).
\eqne
In position space this integral can be expressed by
\eqb\label{exd3p}
D_3(p) = -\frac12\sum_{x\in\Lambda}\mbox{e}^{ipx}\tilde D_3(x),
\eqe
where
\eqb
\tilde D_3(x) =
G(x)^2\sum_{\mu=0}^1\Big\{
\Big[(\partial_{\mu}-\partial_{\mu}^*)G(x)\Big]^2+
\Big[(\partial_{\mu}+\partial_{\mu}^*)G(x)\Big]^2\Big\}.
\eqe

Up to order $|x|^{-4}$, we obtain for the large $x$ expansion of 
the function $\tilde D_3(x)$
\eqb
\tilde D_3(x) = \Big\{\tilde D_3(x)\Big\}_{\infty} + {\cal O}(|x|^{-6}),
\eqe
where
\eqnb
\Big\{\tilde D_3(x)\Big\}_{\infty} &=&
\frac{1}{16\pi^4}\frac{1}{x^2}
\bigg\{(\ln x^2)^2+2Q_0\ln x^2+Q_0^2 \nonumber\\ & &
+\frac{1}{x^2}\bigg[-\frac72(\ln x^2)^2+(1-7Q_0)\ln x^2-\frac72Q_0^2+Q_0\bigg]
\nonumber\\\label{d3expan} & &
+\frac{x^4}{(x^2)^3}\bigg[5(\ln x^2)^2+
(10Q_0-\frac43)\ln x^2+5Q_0^2-\frac43Q_0\bigg]\bigg\}
\eqne
with $Q_0=2\gamma+3\ln 2$.

The sum (\ref{exd3p})
can now be evaluated in exactly the same way as $D_2(p)$.
The only difference in this case is 
that, due to the quadratically logarithmic term
in the expansion (\ref{d3expan}),
it is necessary to differentiate zeta functions two times
with respect \mbox{to $s$.}
After a straightforward computation, we get for $D_3(p)$
\eqb
D_3(p) = \frac{1}{32\pi^3}\bigg[\frac13(\ln p^2)^3
-5\ln 2(\ln p^2)^2+(5\ln 2)^2\ln p^2\bigg] + \kappa_2 
\eqe
with $\kappa_2$ given by
\eqb
\kappa_2 = -0.01802345(1).
\eqe

\section*{Acknoledgements}

I would like to thank Peter Weisz for many fruitful discussions.
I express my thanks also to Andrea Pelissetto who informed me of the analytic
value for $A_1$ of Eq.~(\ref{anal124}).

\begin{appendix}

\section{Large $x$ expansion of $G(x)$}\label{ap1}
\setcounter{equation}{0}

In this appendix we derive for the free propagator $G(x)$
given in Eq.~(\ref{freeprg})
the large~$x$ expansion of Eqs.~(\ref{gxcm}) and (\ref{gxlx}).

As a first step, we use Cauchy's integral formula 
to do the $p_0$ integration and for simplicity
set $x=(x_0,0)$. Then\footnote{For a given function $F(x)$, the
simplified notation $F(x_0)$ stands for the exact $F(x)|_{x=(x_0,0)}$.}
\eqb\label{1dimi}
G(x_0) = \int^{\pi}_{0}\frac{\md p}{\pi}\,
\frac{\me^{x_0\ln\frac12\big[\hat p^2+2 -\sqrt{\hat p^2 (\hat p^2 +4)}
\big]}-1}{\sqrt{\hat p^2 (\hat p^2+4)}} .
\eqe
Change variable $y=\sin p/2$, 
then $\hat p=2y$, $\md p=2\md y/\sqrt{1-y^2}$
and Eq.~(\ref{1dimi}) becomes
\eqb\label{varcha}
G(x_0) = \frac{1}{2\pi}\int^{1}_{0}\frac{\md y}{y}\,
\frac{\me^{x_0f(y)}-1}{\sqrt{1-y^4}} ,
\eqe
where $f(y)$ is defined by
\eqb\label{deffy}
f(y) = \ln \Big(1+2y^2 -2y\sqrt{1+y^2}\Big) .
\eqe
Now, divide the integral of Eq.~(\ref{varcha}) into two parts
\eqb
G(x_0) = \frac{1}{2\pi}\Big[\bar G(x_0) + \tilde G(x_0)\Big],
\eqe
where
\eqnb\label{brx0}
\bar G(x_0) &=&
\int^{1}_{0}\frac{\md y}{y}\,
\bigg[\frac{\me^{x_0f(y)}}{\sqrt{1-y^4}}-1\bigg] , \\\label{tilgx0} 
\tilde G(x_0) &=& 
\int^{1}_{0}\frac{\md y}{y}\,
\bigg(1-\frac{1}{\sqrt{1-y^4}}\bigg) .
\eqne

We are interested in the evaluation of $G(x_0)$ in the limit $x_0\to\infty$. 
For this purpose, we perform another 
variable change $t=x_0y$ in $\bar G(x_0)$.
In the continuum limit ($x_0\to\infty$), Eq.~(\ref{brx0}) then reduces to
\eqb\label{bx0lx0}
\bar G(x_0) \sim\int^{x_0}_{0}\md t\, \frac{\me^{-2t}-1}{t} .
\eqe
The integrals of Eqs.~(\ref{tilgx0}) and (\ref{bx0lx0})
can now be computed analytically and we finally get for
the leading terms in the continuum limit
\eqb\label{gx0cont}
G^{{\rm c}}(x_0) = -\frac{1}{4\pi}\big(2\ln x_0+2\gamma+3\ln 2\big) .
\eqe
Using rotation invariance in the continuum limit, 
the leading behavior for $G(x)$ is then equal to
\eqb\label{gxcont}
G^{{\rm c}}(x) = -\frac{1}{4\pi}(\ln x^2 + 2\gamma+3\ln 2).
\eqe

The subleading terms can also be computed systematically 
if we make use of the laplace equation~(\ref{lapla}).
We derive, at first, the first subleading terms through the ansatz
\eqb\label{suban1}
G(x) = G^{{\rm c}}(x) + G^{(1)}(x) + {\cal O}(|x|^{-4}) 
\eqe
with
\eqb
G^{(1)}(x) = -\frac{1}{4\pi}
\bigg[c_1\frac{1}{x^2} + c_2\frac{x^4}{(x^2)^3}\bigg].
\eqe
For determination of $c_1$ and $c_2$, we evaluate
the lattice laplacian of $G(x)$ in Eq.~(\ref{suban1}) to get
\eqb\label{latlap1}
\triangle G(x) = 
-\frac{1}{4\pi}\bigg\{\frac{6}{(x^2)^2}-8\frac{x^4}{(x^2)^4}
+ c_1\frac{4}{(x^2)^2}
+ 12c_2\bigg[\frac{1}{(x^2)^2}-\frac{x^4}{(x^2)^4}\bigg]\bigg\}
+{\cal O}(|x|^{-6}) .
\eqe
From the condition that the r.h.s. of Eq.~(\ref{latlap1}) should
be zero, we obtain
\eqb
c_1 = \frac12\,\,, \hspace{1cm} c_2 = -\frac23\,\, .
\eqe

Now, we compute the next subleading terms and try with the ansatz
\eqb\label{seclansa}
G(x) = G^{{\rm c}}(x) + G^{(1)}(x) + G^{(2)}(x) + {\cal O}(|x|^{-6}) ,
\eqe
where
\eqb
G^{(2)}(x) = -\frac{1}{4\pi}\bigg[c_3\frac{1}{(x^2)^2}
+ c_4\frac{x^4}{(x^2)^4} + c_5\frac{(x^4)^2}{(x^2)^6}\bigg].
\eqe
The lattice laplacian of this function is given by
\eqnb
\triangle G(x) &=& 
-\frac{1}{4\pi}\bigg\{
-\frac{170}{(x^2)^3}+480\frac{x^4}{(x^2)^5}
-320\frac{(x^4)^2}{(x^2)^7}
+ c_3\frac{16}{(x^2)^3} \nonumber\\ & &
+ c_4\frac{12}{(x^2)^3} 
- c_5\bigg[\frac{16}{(x^2)^3}-72\frac{x^4}{(x^2)^5}+48\frac{(x^4)^2}{(x^2)^7}
\bigg]\bigg\}
+{\cal O}(|x|^{-8}) .
\eqne
Since the r.h.s. has to vanish, we obtain
\eqb\label{1stcon}
24c_3+18c_4-95=0\,\,, \hspace{1cm} c_5 = -\frac{20}{3}\,\,.
\eqe

Unfortunately, the condition from the laplace equation
alone is not enough to determine the coefficients completely
and we need another constraint for them.
It can be obtained by computing
the subleading terms of $G(x_0)$ in Eq.~(\ref{1dimi}) to this order.
After some algebra, we find
\eqb
G(x_0) = G^{{\rm c}}(x_0) + 
\frac{1}{24\pi}\frac{1}{x_0^2}\bigg(1+\frac{43}{20}\frac{1}{x_0^2}\bigg)
+{\cal O}(x_0^{-6}).
\eqe
By comparing this with Eq.~(\ref{seclansa}) for $x=(x_0,0)$, 
we get the relation
\eqb\label{2ndcon}
c_3+c_4+c_5 = -\frac{43}{120}\,.
\eqe
From Eqs.~(\ref{1stcon}) and (\ref{2ndcon}), the coefficients
$c_3$ and $c_4$ can now be determined:
\eqb
c_3 = -\frac{371}{120}\,\,, \hspace{1cm} c_4 = \frac{47}{5}\,\,.
\eqe

The computation of the third subleading terms proceeds completely analogously.
After a lengthy, but straightforward computation, we find 
\eqb
G(x) = G^{{\rm c}}(x) + G^{(1)}(x) + G^{(2)}(x) + G^{(3)}(x) + 
{\cal O}(|x|^{-8}) 
\eqe
with
\eqb
G^{(3)}(x) = -\frac{1}{4\pi}
\bigg[
\frac{4523}{56}\frac{1}{(x^2)^3} - \frac{7657}{21}\frac{x^4}{(x^2)^5} 
+ \frac{3716}{7}\frac{(x^4)^2}{(x^2)^7} 
- \frac{2240}{9}\frac{(x^4)^3}{(x^2)^9}\bigg].
\eqe

\section{Large $x$ expansion of $G_2(x)$}\label{ap2}
\setcounter{equation}{0}

We derive here the large $x$ expansion of Eq.~(\ref{laxg2}) for the function
$G_2(x)$ given in Eq.~(\ref{defg2}).

We first do the $p_0$ integration and for simplicity
set $x=(x_0,0)$. Then 
\eqnb
G_2(x_0) &=& \int^{\pi}_{0}\frac{\md p}{\pi}\,
\bigg\{\frac{(x_0+1)\me^{x_0\ln\frac12 (\hat p^2+2-P)}
-1-\frac{x_0^2}{2}(\hat p^2-P)}{P^2} \nonumber\\\label{g2p0int} & & 
+\frac{(\hat p^2+2-P)
\big[\me^{x_0\ln\frac12(\hat p^2+2-P)}-1\big]
-\frac{x_0^2}{4}(\hat p^2-P)^2}{P^3}\bigg\}
\eqne
with $P=\sqrt{\hat p^2(\hat p^2+4)}$.
By changing variable $y=\sin p/2$, Eq.~(\ref{g2p0int}) transforms to
\eqb\label{g2x0y}
G_2(x_0) = \frac{1}{16\pi}\int^{1}_{0}\frac{\md y}{y^3}\,
\frac{\me^{x_0f(y)}(2y^2+2x_0y\sqrt{1+y^2}+1)-[1+2(1-x_0^2)y^2]}
{(1+y^2)\sqrt{1-y^4}}
\eqe
with $f(y)$ given in Eq.~(\ref{deffy}).

Now, divide the integral of Eq.~(\ref{g2x0y}) into two parts:
\eqb
G_2(x_0) = \frac{1}{16\pi}\Big[\bar G_2(x_0) + \tilde G_2(x_0)\Big],
\eqe
where $\big[Y=(1+y^2)\sqrt{1-y^4}\big]$
\eqnb\label{g2x02p}
\bar G_2(x_0) &=& \int^{1}_{0}\frac{\md y}{y^3}\,
\bigg\{\frac{\me^{x_0f(y)}(2y^2+2x_0y\sqrt{1+y^2}+1)}{Y}
-\Big[1+\big(1-2x_0^2\big)y^2\Big]\bigg\}, \\ 
\tilde G_2(x_0) &=& \int^{1}_{0}\frac{\md y}{y^3}\,
\bigg\{1+y^2-\frac{1+2y^2}{Y}\bigg\} 
-2x_0^2\int^{1}_{0}\frac{\md y}{y}\,
\bigg(1-\frac{1}{Y}\bigg) .
\eqne

We are interested in the evaluation of $G_2(x_0)$ 
in the continuum limit $x_0\to\infty$. 
For this purpose, we perform the variable change $t=x_0y$.
In the continuum limit, Eq.~(\ref{g2x02p}) then reduces to
\eqnb
\bar G_2(x_0) &\sim & \int^{x_0}_{0}\frac{\md t}{t^3}\,
\Big\{\me^{-2t}[(1+2t)x_0^2+t^2-t^3]-(1-2t^2)x_0^2-t^2\Big\}\nonumber\\ & &
+\frac13\int^{x_0}_{0}\md t\,
\me^{-2t}(1+2t) .
\eqne
In the limit of infinitely large $x_0$, 
these integrals can be evaluated analytically to yield
\eqb
\bar G_2(x_0) \sim x_0^2\,(2\ln{x_0}+2\ln 2+2\gamma-1)-\ln x_0-\ln 2-\gamma
+\frac13 \, .
\eqe
As for $\tilde G_2(x_0)$, we can also compute the function analytically:
\eqb
\tilde G_2(x_0) = x_0^2\,(\ln 2-1)-\frac12\ln 2\, .
\eqe
Summing two contributions together, we have finally
for the leading terms in the continuum limit
\eqb\label{resg2x0}
G_2^{{\rm c}}(x_0) = \frac{1}{16\pi}
\Big[x_0^2\,(2\ln{x_0}+3\ln 2+2\gamma -2)-\ln x_0-\frac32\ln 2-\gamma
+\frac13\Big].
\eqe

Unfortunately, from Eq.~(\ref{resg2x0}) alone 
one can not derive the corresponding function
$G_2^{{\rm c}}(x)$ unambiguously
since $G_2(x)$ in the continuum limit is not rotation invariant.
We therefore apply the laplace equation (\ref{lapg2}) 
from which we obtain for $G_2(x)$
the following leading terms in the infinitely large $x$ limit: 
\eqb
G_2^{{\rm c}}(x) = \frac{1}{16\pi}
\Big[x^2\ln x^2+(3\ln 2+2\gamma -2)x^2+\kappa_1\ln x^2+
\frac13\frac{x^4}{(x^2)^2}+\kappa_2\Big] .
\eqe
Comparison of this result for $x=(x_0,0)$
with Eq.~(\ref{resg2x0}) gives finally
\eqb
\kappa_1 = -\frac12\,\,, \hspace{1cm} 
\kappa_2 = -\frac12(3\ln 2 +2\gamma)\,\, .
\eqe

The subleading terms for $G_2(x)$
can also be worked out systematically
by applying the laplace equation (\ref{lapg2}) 
and the asymptotic expansion for $G_2(x_0)$ of Eq.~(\ref{g2p0int}).
We do not present here the detailed computations leading to the result of 
Eq.~(\ref{laxg2}) since they are too long and also proceed 
very similarly to the case of $G(x)$.

\section{Derivation of harmonic functions}\label{aphhp}
\setcounter{equation}{0}

We discuss here how one computes the harmonic homogeneous polynomials 
appearing in the main text.

Let us consider the following polynomial functions of degree $4n$
\eqb\label{hnxpol}
h_n(x) = \sum_{r=0}^nC_r^{(n)}(x^4)^r(x^2)^{2(n-r)}
\eqe
with $C_n^{(n)}\not=0$.
We would like to derive the conditions for the coefficients $C_r^{(n)}$
so that $h_n(x)$ are harmonic functions.

For determination of the conditions, we first note the relation
\eqnb
\lefteqn{
\triangle_{{\rm c}}\Big[(x^4)^r(x^2)^{s}\Big] =
4\Big[
-2r(r-1)(x^4)^{r-2}(x^2)^{s+3}}
\nonumber\\\label{laplhnx} & &
+3r(2r-1)(x^4)^{r-1}(x^2)^{s+1}
+s(4r+s)(x^4)^{r}(x^2)^{s-1}\Big],
\eqne
where $\triangle_{{\rm c}}$ denotes the continuum laplacian
$\sum_{\mu=0}^1\partial^2/\partial x_{\mu}^2$.
In deriving the relation above we made use of the identity~(\ref{identx6}).
By using Eq.~(\ref{laplhnx}) we obtain
\eqnb
\triangle_{{\rm c}} h_n(x) &=&
4\sum_{r=0}^n\Big[
4C_{r}^{(n)}(n^2-r^2)
+3C_{r+1}^{(n)}(r+1)(2r+1)\theta(n-1-r) \nonumber\\\label{laplhtr} & & 
-2C_{r+2}^{(n)}(r+1)(r+2)\theta(n-2-r) 
\Big](x^4)^{r}(x^2)^{2n-2r-1} \,,
\eqne
where $\theta(t)$ is a step function defined by
\begin{displaymath}
\theta(t) =
\left\{\begin{array}{ll}
1\hspace{0.5cm}\mbox{for}\hspace{0.5cm}t\geq 0\,, \\
0\hspace{0.5cm}\mbox{for}\hspace{0.5cm}t< 0\,.
\end{array} \right.
\end{displaymath}

For $h_n(x)$ to be harmonic, the r.h.s. of Eq.~(\ref{laplhtr}) 
should be zero, which
leads to the conditions for the coefficients $C_r^{(n)}$ to satisfy.
Now, we investigate these conditions by dividing the $n+1$ terms
in the summation
contributing to $\triangle_{{\rm c}} h_n(x)$ into the following
three cases:
\begin{enumerate}
\item For $r=n$: the contribution to $\triangle_{{\rm c}} h_n(x)$ is
identically zero and there is therefore no constraint on the 
coefficients $C_r^{(n)}$.
\item For $r=n-1$: the contribution to $\triangle_{{\rm c}} h_n(x)$ is zero if
\eqb\label{harcon1}
3nC_n^{(n)}+4C_{n-1}^{(n)}=0\,.
\eqe
\item For $r\leq n-2$: 
the contribution to $\triangle_{{\rm c}} h_n(x)$ is zero if
\eqb\label{harcon2}
4C_{r}^{(n)}(n^2-r^2)+3C_{r+1}^{(n)}(r+1)(2r+1)-2C_{r+2}^{(n)}(r+1)(r+2) = 0\,.
\eqe
\end{enumerate}

It is now straightforward to derive the 
harmonic homogeneous polynomials including those of
Eqs.~(\ref{harmh0})--(\ref{harmh2}) by solving the recursion relations
(\ref{harcon1}) and (\ref{harcon2}) up to an arbitrary overall 
normalization factor.

\section{Analytic computation of $A_1$}\label{ap3}
\setcounter{equation}{0}

This appendix where we evaluate the two-loop integral $A_1$ given in 
Eq.~(\ref{anal124}) analytically
is due to communication with A. Pelissetto.

We first use the identity $\hat s_{\mu}\cos\frac{s_{\mu}}{2}=\sin s_{\mu}$
and the permutation symmetry bet\-ween \mbox{$k$, $l$} and $(-s)$
to transform the given integral to
\eqb
A_1 = \frac13\int^{\pi}_{-\pi}
\frac{\md^2k}{(2\pi)^2}\frac{\md^2l}{(2\pi)^2}
\frac{1}{\hat k^2\hat l^2\hat{s}^2}
\sum_{\mu=0}^1\hat k_{\mu}\hat l_{\mu}\hat s_{\mu}
(\sin s_{\mu}-\sin k_{\mu}-\sin l_{\mu}) .
\eqe
Now, we use the following
trigonometric identity being valid for $\alpha+\beta+\gamma=0$
\eqb
\sin\alpha+\sin\beta+\sin\gamma = 
-4\sin\frac{\alpha}{2}\sin\frac{\beta}{2}\sin\frac{\gamma}{2} 
\eqe
to rewrite
\eqb
A_1 = -\frac43\int^{\pi}_{-\pi}
\frac{\md^2k}{(2\pi)^2}\frac{\md^2l}{(2\pi)^2}
\frac{1}{\hat k^2\hat l^2\hat{s}^2}
\sum_{\mu=0}^1\hat k_{\mu}\hat l_{\mu}\hat s_{\mu}
\sin\frac{k_{\mu}}{2}\sin\frac{l_{\mu}}{2}\sin\frac{s_{\mu}}{2}.
\eqe
Noting the relation $\sin\frac{s_{\mu}}{2}=\frac12\hat s_{\mu}$, we see that 
this integral is, apart from a factor, same as $A^{(3)}$ in Appendix~A
of Ref.~\cite{CaPeII} which is known analytically.
From this, we obtain
\eqb
A_1 = -\frac{1}{24}.
\eqe

\end{appendix}

\end{document}